\documentclass[aps,prl,reprint,showpacs,superscriptaddress,twocolumn]{revtex4}
\usepackage[dvips]{graphicx}
\usepackage{subfigure}
\usepackage{bm}
\usepackage{color}

\begin{document}

\title{
Continuous Generation of Spinmotive Force in a Patterned Ferromagnetic Film
}

\author{Y. Yamane}%$^{\ast\ 1,2}$
\email[equal contribution: ]{yutaymn322@gmail.com}
\affiliation{Advanced Science Research Center, Japan Atomic Energy Agency, Tokai 319-1195, Japan}
\affiliation{Institute for Materials Research, Tohoku University, Sendai 980-8577, Japan}
\author{K. Sasage}%$^{\ast\ 2}$
\email[equal contribution: ]{k.sasage@imr.tohoku.ac.jp}
\affiliation{Institute for Materials Research, Tohoku University, Sendai 980-8577, Japan}
\affiliation{CREST, Japan Science and Technology Agency, Tokyo 102-0075, Japan}
\author{T. An}%$^2$,
\affiliation{Institute for Materials Research, Tohoku University, Sendai 980-8577, Japan}
\affiliation{CREST, Japan Science and Technology Agency, Tokyo 102-0075, Japan}
\author{K. Harii}%$^2$,
\affiliation{Institute for Materials Research, Tohoku University, Sendai 980-8577, Japan}
\affiliation{CREST, Japan Science and Technology Agency, Tokyo 102-0075, Japan}
\author{J. Ohe}%$^{3,4}$,
\affiliation{CREST, Japan Science and Technology Agency, Tokyo 102-0075, Japan}
\affiliation{Department of Physics, Toho University, Funabashi, 274-8510 Japan}
\author{J. Ieda}%$^{1,3}$,
\affiliation{Advanced Science Research Center, Japan Atomic Energy Agency, Tokai 319-1195, Japan}
\affiliation{CREST, Japan Science and Technology Agency, Tokyo 102-0075, Japan}
\author{S. E. Barnes}%$^5$,
\affiliation{Physics Department, University of Miami, Coral Gables, Florida 33124, USA}
\author{E. Saitoh}%$^{1,2,3}$,
\affiliation{Advanced Science Research Center, Japan Atomic Energy Agency, Tokai 319-1195, Japan}
\affiliation{Institute for Materials Research, Tohoku University, Sendai 980-8577, Japan}
\affiliation{CREST, Japan Science and Technology Agency, Tokyo 102-0075, Japan}
\author{S. Maekawa}%$^{1,3}$}
\affiliation{Advanced Science Research Center, Japan Atomic Energy Agency, Tokai 319-1195, Japan}
\affiliation{CREST, Japan Science and Technology Agency, Tokyo 102-0075, Japan}

\date{\today}

\begin{abstract}
We study, both experimentally and theoretically, the generation of a dc spinmotive force.
By exciting a ferromagnetic resonance of a comb-shaped ferromagnetic thin film,  a continuous spinmotive force is generated.
Experimental results are well reproduced by theoretical calculations, offering a quantitative and microscopic understanding of this spinmotive force.
\end{abstract}

% Barnes Maekawa PRL 2007 Generalized Faraday Law...
% 72.25.Ba	Spin polarized transport in metals
% 03.65.Vf	Phases: geometric; dynamic or topological
% 75.60.Ch	Domain walls and domain structure
% 85.75.-d	Magnetoelectronics; spintronics: devices exploiting spin polarized transport or integrated magnetic fields

% 75.78.Cd 	Micromagnetic simulations
% 76.50.+g 	Ferromagnetic, antiferromagnetic, and ferrimagnetic resonances; spin-wave resonance

\maketitle

%===================================================================================================
%           Introduction
%===================================================================================================
%\footnotemark[0]
\emph{Introduction.}---
In classical electrodynamics, Faraday's law equates the electromotive force to the time derivative of a magnetic flux.
Involved is a coupling to the electrical charge of electrons.
Recently, a motive force of spin origin, i.e., a ``spinmotive force", has been theoretically predicted\cite{barnes1} and experimentally observed\cite{yang,tanaka}.
A spinmotive force reflects the conversion of the magnetic energy of a ferromagnet into the electrical energy of the conduction electrons this via their mutual exchange interaction.
A spinmotive force reflects the spin  of electrons in an essential manner.
It is a new concept  relevant to electronic devices\cite{barnes2}.

However, problems remain, one of which is the continuous generation of such a  spinmotive force.
Theoretically, it has been pointed out that the spinmotive force is produced by the spin electric field\cite{field}
\begin{equation}
\bm{E}_\mathrm{s}=-\frac{P\hbar}{2e}\bm{m}\cdot\left(\partial_t\bm{m}\times\nabla\bm{m}\right)
\label{e}\end{equation}
where $\bm{m}$ is the unit vector parallel to the  local magnetization direction of the ferromagnet, $P$ the spin polarization of the conduction electrons, and $e$ the elementary charge.
It is required that the magnetization depends both on the time and space.
For recent experiments involving domain wall motion\cite{yang}, or electron transport through ferromagnetic nanoparticles\cite{tanaka}, the magnetization motion and hence the spinmotive force are, by their nature, transient.
On the basis of numerical work\cite{ohe}, it is predicated that a continuous ac spinmotive force is predicted to accompany the gyration motion of a magnetic vortex core.

In this Letter, we present experiment and theory that demonstrate the continuous generation of a dc spinmotive force.
The spin electric field Eq.~(\ref{e}) can be induced by exciting the ferromagnetic resonance (FMR) of a highly asymmetrical ferromagnetic thin film.
It has already been shown that an electrical voltage can be generated in a ferromagnet(F)/normal-metal(N) junction excited by FMR.
This is interpreted in terms of the spin accumulation at the interface\cite{pump1}.
As will be discussed in more detail below, here a voltage generation uses a single ferromagnet for which no appreciable spin accumulation is possible and for which this scenario is ruled out.
The observed electromotive force is well reproduced by simulations based on the spinmotive force theory.
 \begin{figure}[b]
     \begin{center}
         \includegraphics[width=3.7cm]{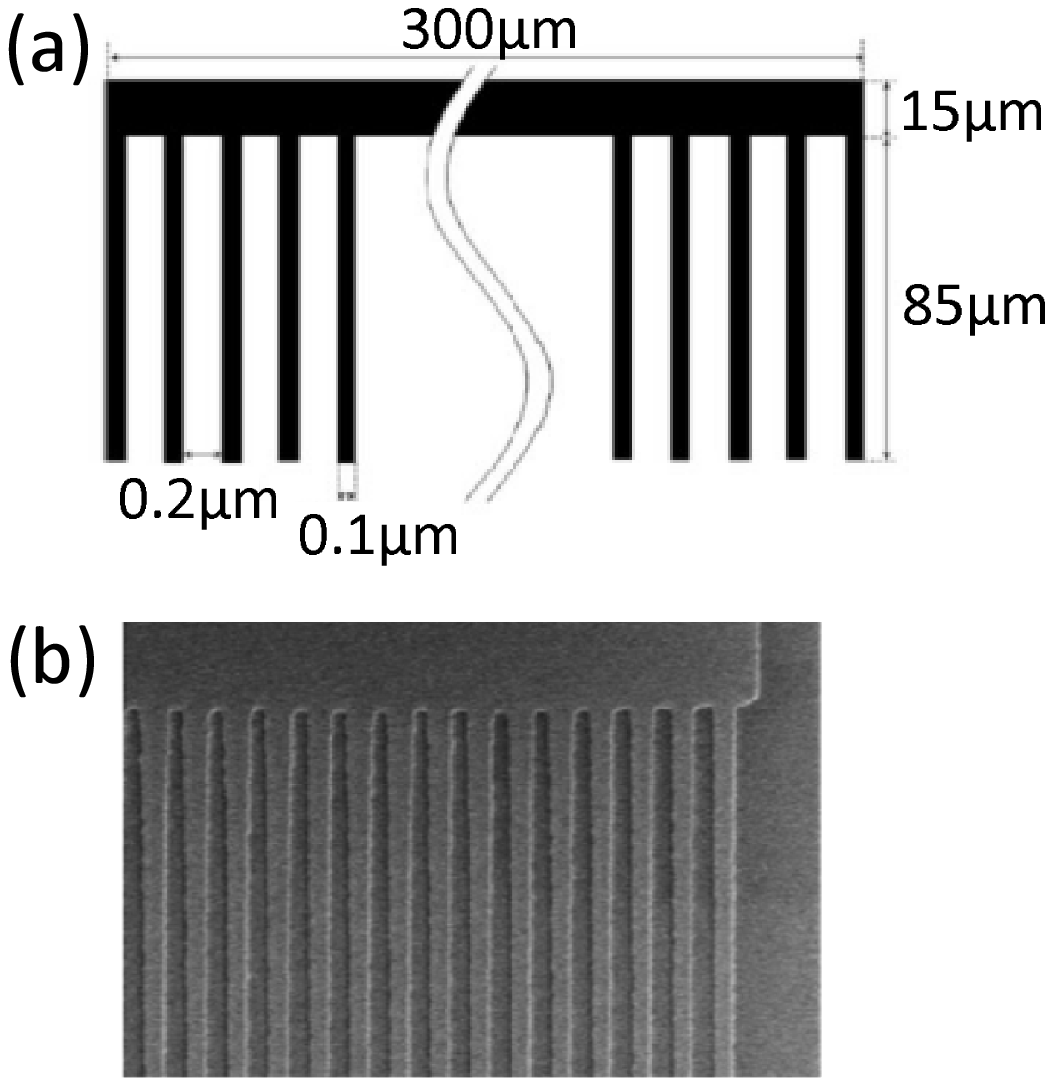}
         \includegraphics[width=4.8cm]{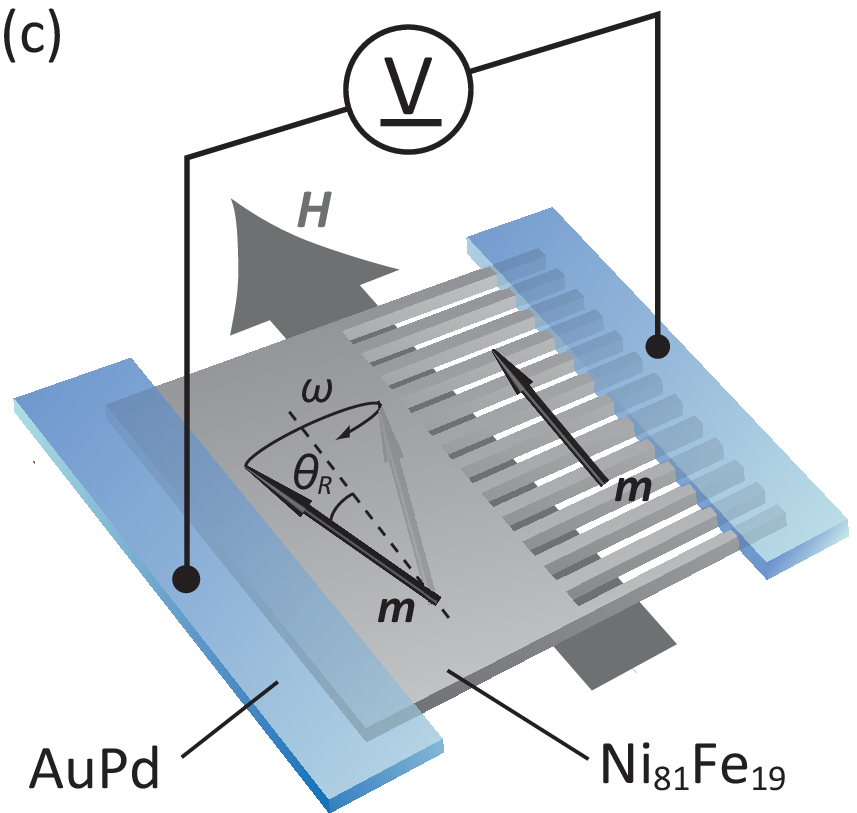}
         \caption{  (a) A schematic illustration of the Ni$_{81}$Fe$_{19}$ ``comb" sample of thickness is 20 nm, and composed of a large pad and an array of wires.
                    (b) A SEM image of the junction region between the pad and wires.
                    (c) A cartoon of the experimental procedure.
                        The local magnetization $\bm{m}$ and the applied static magnetic field $\bm{H}$ are shown. 
                        The pad FMR is excited without that of the wires due to different shape anisotropies. (The inverse also occurs.)
                        In a macro magnetic moments picture, the resonant magnetization is characterized by its precession angle $\theta_R$ and angular frequency $\omega$.
                        A dc electromotive force is measured by two normal metal electrodes, Au$_{25}$Pd$_{75}$, shown as blue rectangles.
                 }
         \label{fig1}
     \end{center}
 \end{figure}

%===================================================================================================
%           Experiment
%===================================================================================================
\emph{Experiment.}---
Figure~\ref{fig1} (a) shows a schematic illustration of the experimental sample, a Ni$_{81}$Fe$_{19}$ microstructure ``comb" composed of a large pad connected to an array of wires.
All the relevant  dimensions are shown in Fig. \ref{fig1} (a) and the caption.
The sample was fabricated on a SiO$_2$ substrate by using the electron beam lithography.
A scanning electron microscope (SEM) image of the microstructure is shown in Fig.~\ref{fig1} (b).
In order to reduce the usual induced electromotive force, and as shown schematically in Fig.~\ref{fig1} (c), the system is connected to a voltmeter via two Au$_{25}$Pd$_{75}$ electrodes and a twisted wire pair.
%\subsection{fig2}
\begin{figure}[t]
    \begin{center}
        \begin{tabular}{cc}
            \resizebox{62mm}{!}{\includegraphics{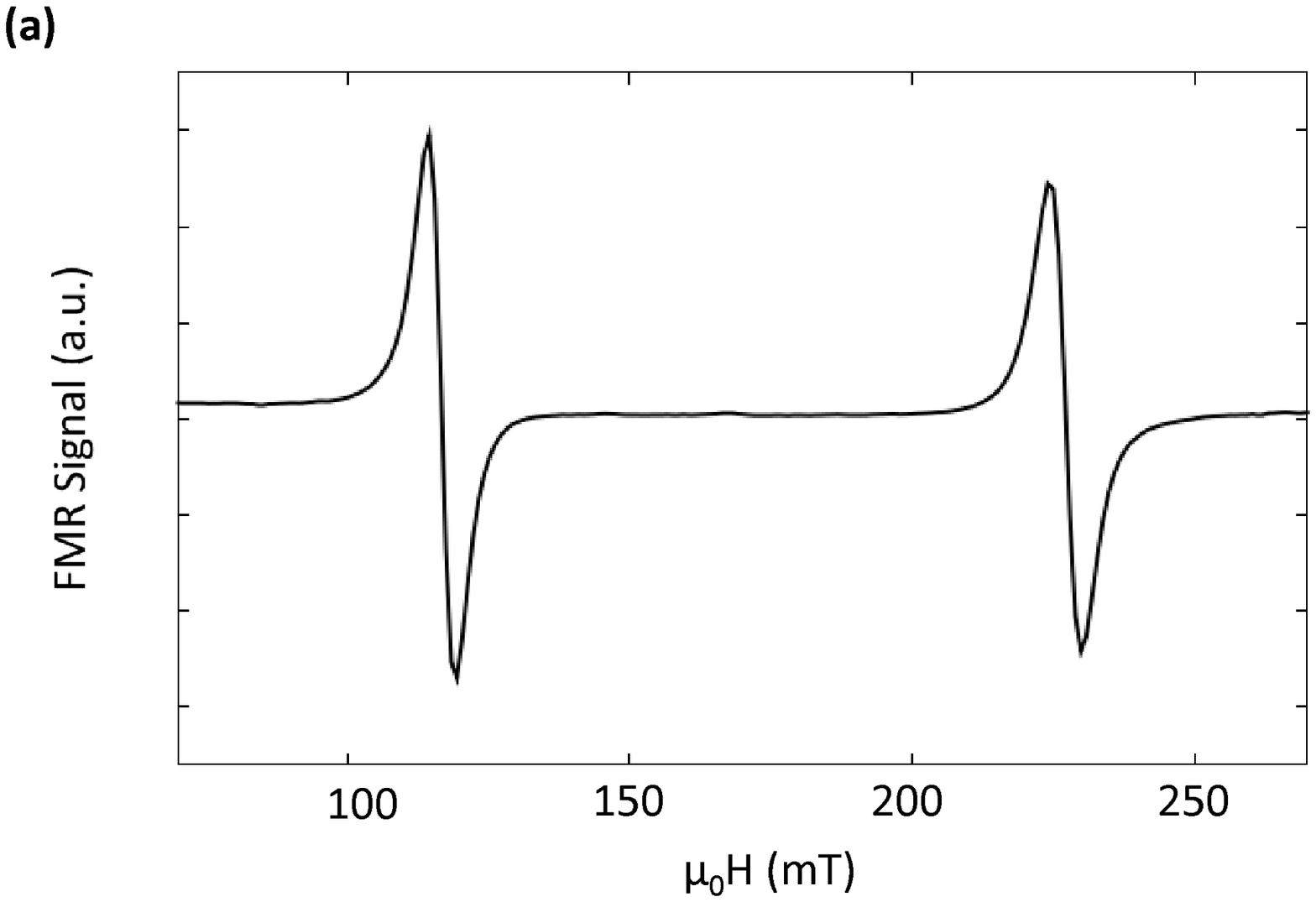}}\\
            \resizebox{62mm}{!}{\includegraphics{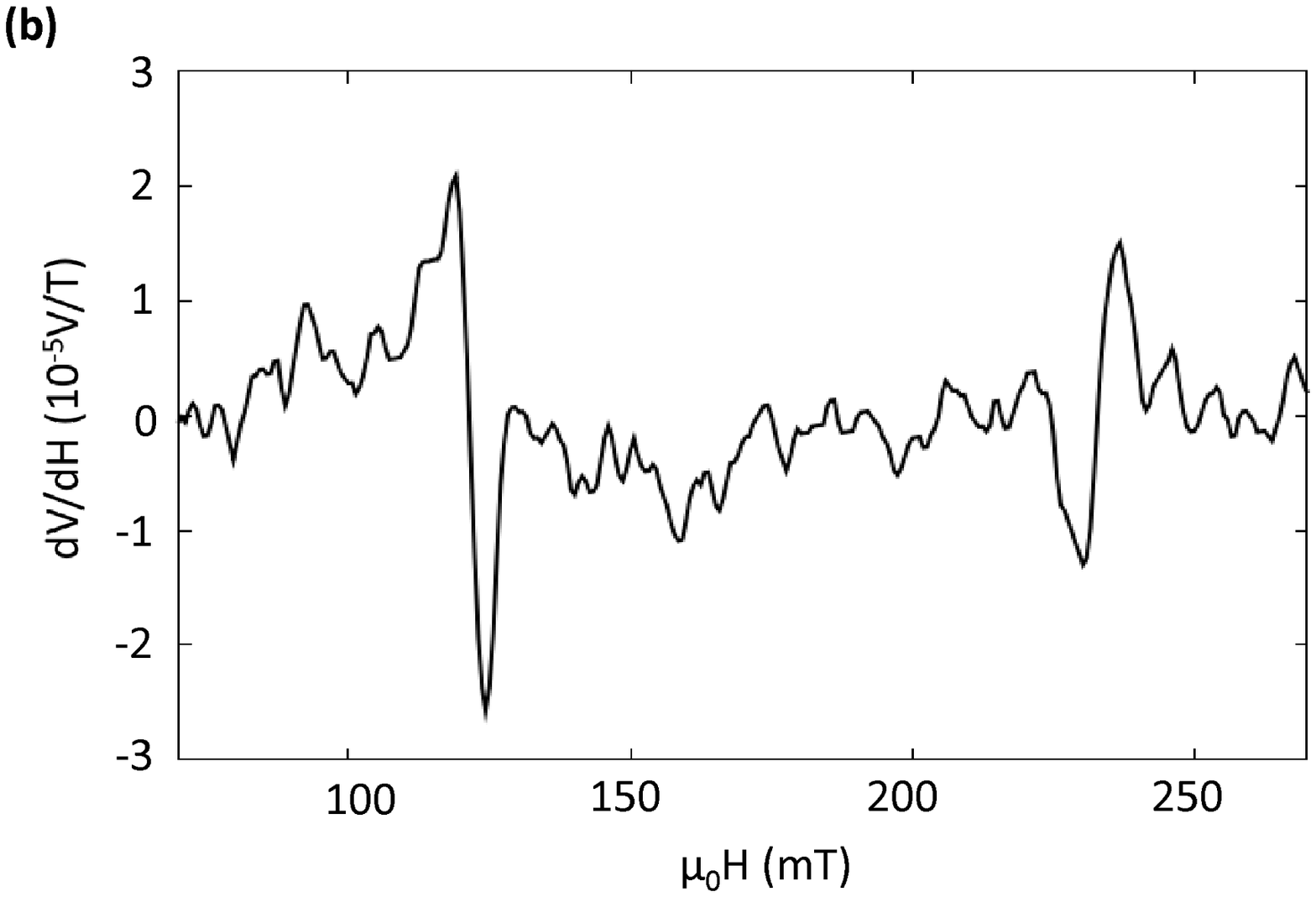}}\\
        \end{tabular}
        \caption{ (a) FMR and (b) electromotive force signals with $200$ mW rf power.
                  The pad ($\mu_0H\approx 120$ mT) and wire ($\mu_0H\approx 230$ mT) FMR signals have the opposite sense only in (b). 
                }
        \label{fig2}
    \end{center}
\end{figure}

Using variable power, a microwave mode, with the frequency of $9.43$ GHz, is excited in a cavity.
An external static magnetic field $H$ is applied parallel to the pad and perpendicular to the wires, see Fig.~\ref{fig2} (c).
The measurements are made at room temperature using lock-in detection.
The static magnetic field modulation at $100$ kHz is $2$ mT, small enough compared with the FMR linewidth of $8$ mT.
Due to the difference in shape anisotropy, the FMR of the pad and the wires can be excited independently, i.e., as shown schematically in Fig.~\ref{fig1} (c), for a static field corresponding to the resonant condition of the pad, the wires remain in equilibrium, and vise versa.
The magnetization configuration is thereby dependent both on the time and space, i.e., the conditions for the appearance of the spin electric field of Eq.~(\ref{e}) are fulfilled.
It is very important to appreciate that the spatial dependence of $\bm{m}$ and hence $\bm{E}_\mathrm{s}$ is localized in the junction region between the pad and the wires that is distant from the contacts, i.e., that the nature of the contacts is not of real importance.
Clearly when one of the pad, or wire, FMR is excited, a spinmotive force will be generated continuously.

\begin{figure}[t]
    \centerline{\includegraphics[width=80mm]{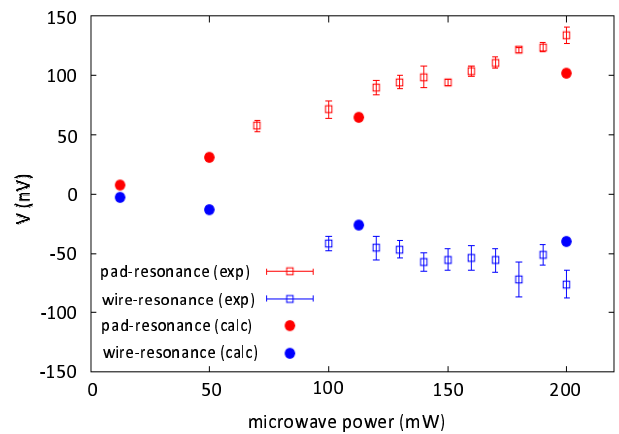}}
    \caption{  Microwave power dependence of the  electromotive force.
               For the pad, experiment and calculation, with $\alpha=0.003$, correspond to red open and closed symbols, respectively.
               Similarly for the wire are blue symbols.
               The magnitude is roughly proportional to the microwave power.
            }
    \label{fig3}
\end{figure}

Figures~\ref{fig2} (a) and (b) show, respectively, the simultaneous  microwave absorption and electromotive force derivative signals for a microwave power of $200$ mW.
Two peaks appear both in the absorption and the electromotive force around $\mu_0H = 120$ mT and $230$ mT, this reflecting, respectively, the resonances of the pad and the wires.
Notice that while the FMR signals have the same polarity the electromotive force equivalents have different signs.
In Fig.~\ref{fig3} is compared the microwave power dependence of the electromotive force, obtained by the integration of the Fig.~\ref{fig2} (b) derivative signal, with that given by theory.
The agreement is satisfactory. 
%\subsection{fig3}

\begin{figure}[t]
    \begin{center}
        \begin{tabular}{cc}
            \resizebox{85mm}{!}{\includegraphics{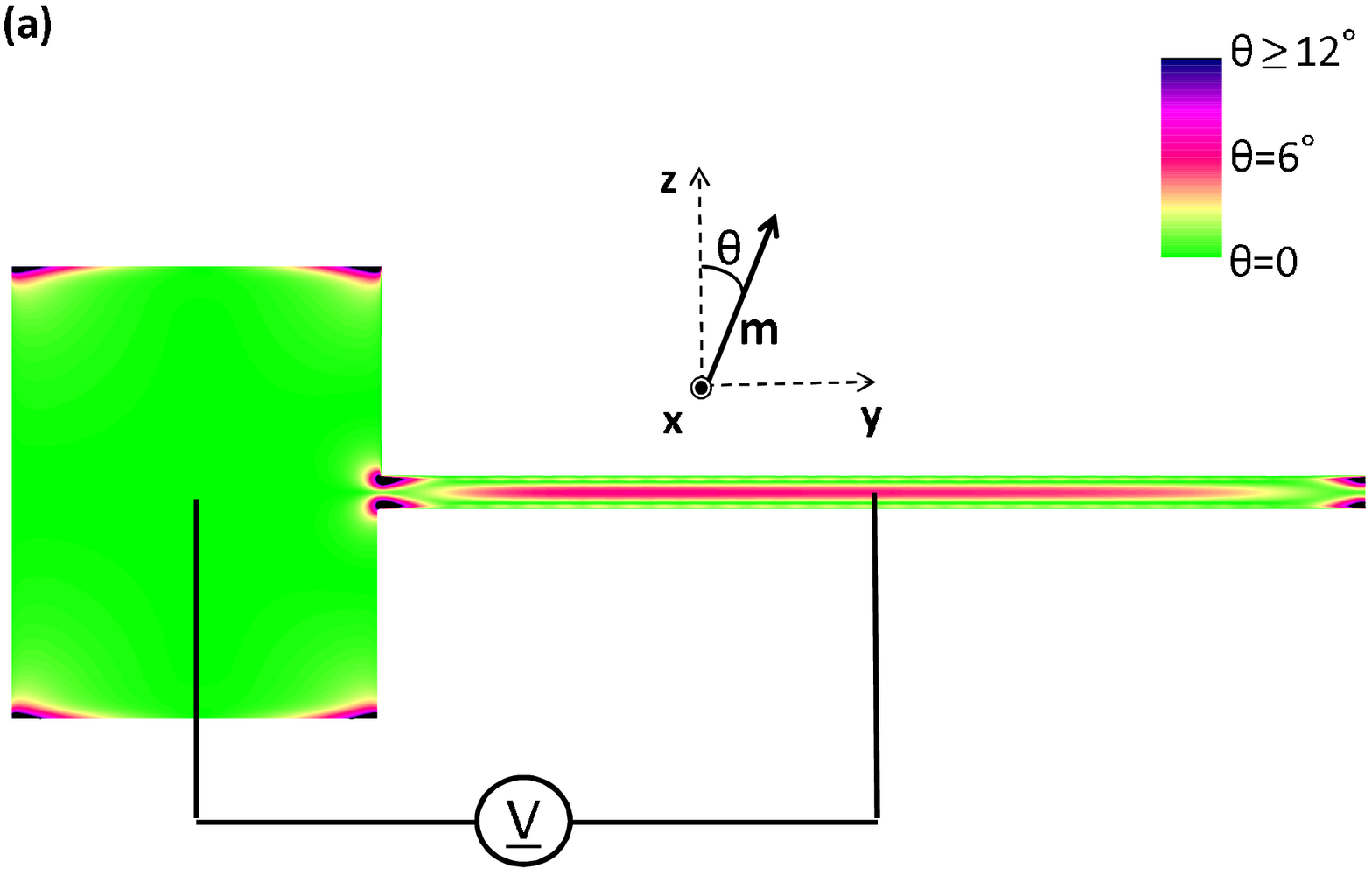}}\\ 
            \resizebox{85mm}{!}{\includegraphics{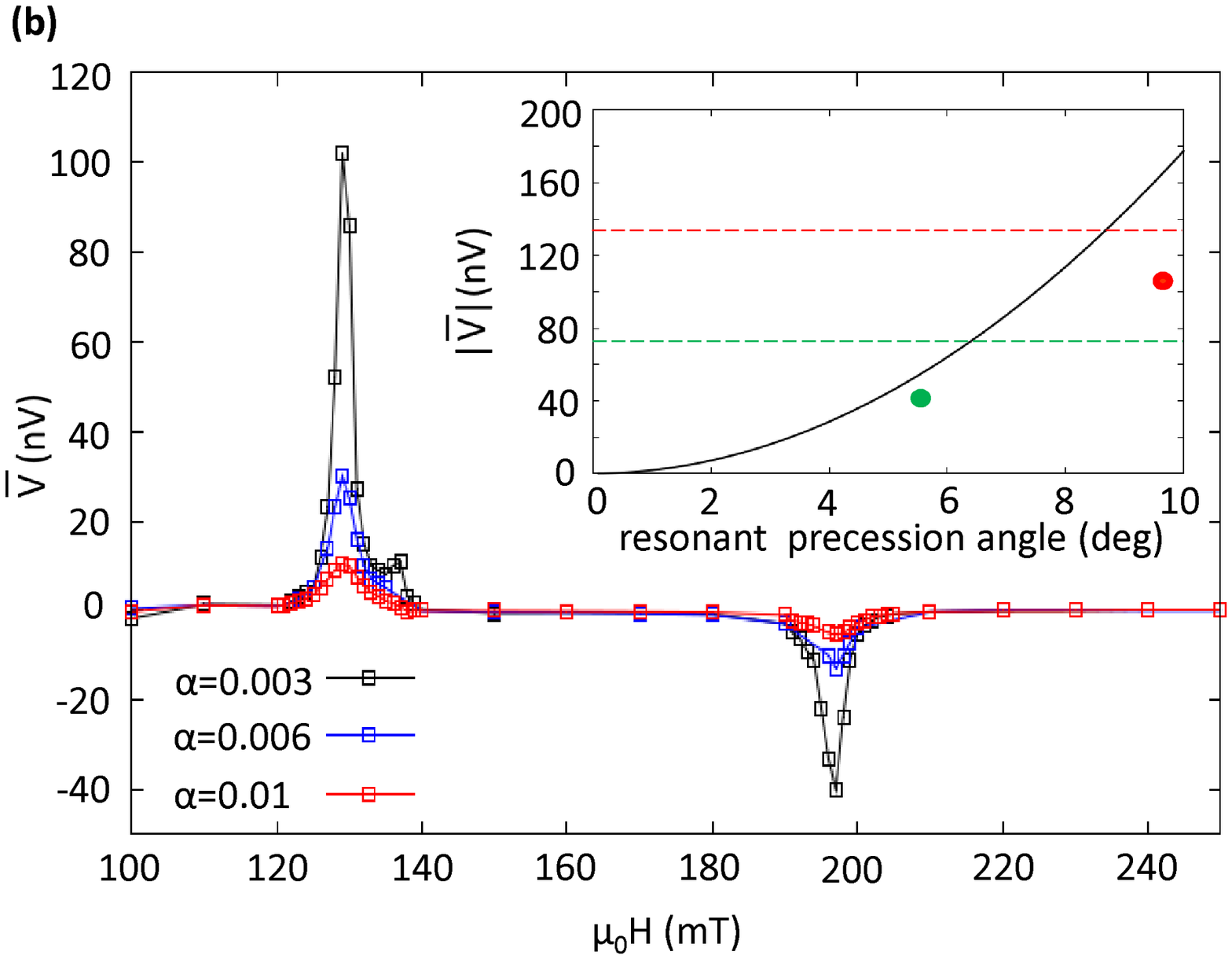}} \\
        \end{tabular}
        \caption{ (a) The shape used in the numerical calculation, which consists of a pad and a single wire.
                      The color indicates the real-space value of the averaged precession angle $\bar{\theta}$ for the wire resonance and $\alpha=0.003$.
                  (b) Calculated average electromotive forces $\bar{V}$ as a function of the  magnetic field $\mu_0H$.
                      Red, blue, and black symbols represent $\bar{V}$ for $\alpha=0.01$, $0.006$ and $0.003$, respectively.
                      Solid lines are guides for the eye.
                      In the inset, the grey solid curve is the spinmotive force Eq.~(\ref{v}) as a function of resonant precession angle $\theta_R$, for the relevant values $P=0.6$ and $\omega=9.43$ GHz.
                      The red (green) dotted line indicates the absolute value of the experimentally observed spinmotive force for the pad- (wire-)resonance.
                      The red (green) circle represents the numerically calculated absolute value of the spinmotive force, versus resonant precession angle, for the pad- (wire-) resonance and $\alpha=0.003$.
                }
        \label{fig4}
    \end{center}
\end{figure}

%===================================================================================================
%           Theoretical study
%===================================================================================================
\emph{Theoretical study.}---
The experimental results can be understood semi-quantitatively by the spinmotive force scenario based on a simple analytical model.
Using $ \bm{m}  =  (  \sin\theta \cos\varphi, \sin\theta \sin\varphi, \cos\theta  ) $, Eq.~(\ref{e}) is re-written as $\textrm{\boldmath $E$}_\mathrm{s}=-\left(P\hbar/2e\right)\sin\theta\left(\partial_t\theta\nabla\varphi-\nabla\theta\partial_t\varphi\right)$.
Consider the situation of Fig.~\ref{fig1} (c) using a macro magnetic moments model for each of the pad and wires, i.e., assume a single uniform $\theta_P$ and $\varphi=\omega t$ for the resonating pad but  $\theta_W\sim0$ (but still $\varphi=\omega t$) for the wires. 
With some sign convention, the electromotive force generated in the region where the pad and wires meet is 
\begin{equation}
%\left|V\right|
V  =  -  \int E_\mathrm{s}dx
   =  \frac{P\hbar\omega}{2e}\left(\cos\theta_W-\cos\theta_P\right) 
\label{v}
\end{equation}
which is approximately $\frac{P\hbar\omega}{2e}\left(1-\cos\theta_P\right)$.
However, when it is rather the wires that resonant, this becomes $\frac{P\hbar\omega}{2e}\left(\cos\theta_W-1\right) $.
Thus Eq.~(\ref{v}) embodies the essential features of the observations:
i) The magnitude of the spinmotive force is determined by the appropriate resonant precession angle $\theta_R = \theta_P$, or $\theta_W$ indicating that the experimental difference in the voltage signals can  attributed to a difference between $\theta_P$ and $\theta_W$.
This is illustrated by the solid line, of the inset of the Fig.~\ref{fig4} (b) which compares $\frac{P\hbar\omega}{2e}\left|1-\cos\theta_R\right| $ with the experimental magnitudes. 
ii) The sign of the spinmotive force is reversed for the pad- and wire-resonance, see Fig.~\ref{fig2} (b), corresponding to the difference in sign between $\left(1-\cos\theta_P\right) $ and $\left(\cos\theta_W-1\right) $. 
iii) When $\theta_R$ is small enough $\left|V\right|\propto{\theta_R}^2$, which is proportional to the microwave power (see Fig.~\ref{fig3}).

However, in reality, even when only one of the pad or wires is excited, the resonant angle $\theta_R$ has a significant spatial dependence.
%\subsection{fig3}
We have therefore carried out numerical calculations of the spin electric field Eq.~(\ref{e}) and resulting spinmotive force for the geometry shown in Fig.~\ref{fig4} (a).
For parameters appropriate to Ni$_{81}$Fe$_{19}$ thin film, considered is a $1.5\times 1.5\times 0.02$ $\mu$m$^3$ pad connected to a single $4\times 0.1\times 0.02$ $\mu$m$^3$ wire.
Since they are electrically in parallel, the number of wires is not directly relevant for the motive force, however the dipole fields do depend on their number and this results in a shift in the wire resonance as compared to the real system.
Also, the number of wires {\it is\/} relevant to the microwave absorption since their signals add.
It is for this reason the wire and pad signals of Fig.~\ref{fig2} (a) are comparable.

To calculate the spin electric field Eq.~(\ref{e}), the dynamics of the magnetization $\bm{m}$ has to be determined.
This is taken to obey the Landau-Lifshitz-Gilbert (LLG) equation,
\begin{equation}
\partial_t\bm{m}=-\gamma\bm{m}\times\textrm{\boldmath $H$}_\mathrm{eff}+\alpha\bm{m}\times\partial_t\bm{m}
\label{llg}\end{equation}
where $\gamma$ the gyromagnetic ratio, $\alpha$ the phenomenological Gilbert damping coefficient, and $\bm{H}_\mathrm{eff}$ is the effective magnetic field.
We calculate numerically the time evolution of Eq.~(\ref{llg}) every $0.01$ ps using the OOMMF code\cite{oommf} with a cell size $10\times 10\times 20$ nm$^3$.
The material parameters are $\gamma=1.76\times 10^{11}$ Hz/T, the saturation magnetization is $740$ mT, $P=0.6$ and the exchange stiffness is $10^{-11}$ A/m.
The phenomenological parameter $\alpha$ is commonly determined from the FMR linewidth; 
in the present experiment, Fig.~\ref{fig2} (a) gives the estimation of $\alpha=0.013$.
However, a FMR linewidth contains effects that are not intrinsic to the material, e.g., the effect of inhomogeneity, which do not affect the local precession of the magnetization.
The {\it intrinsic\/} damping parameter of Ni$_{81}$Fe$_{19}$ is in the range $0.002-0.007$\cite{fmr}.
In order to investigate the dependence of the spinmotive force on $\alpha$, here calculations are performed with three different values of $\alpha=0.01$, $0.006$ and $0.003$.

The assumed external magnetic field is $\textrm{\boldmath $H$}_\mathrm{ext}=\left(h\sin 2\pi ft, 0, H\right)$, with the coordinates shown in Fig.~\ref{fig4} (a), and where $\mu_0h=0.16$ mT, corresponding to $200$ mW, $f=9.43$ GHz and $\mu_0H$ is varied from $100$ to $250$ mT.
The FMR is excited in the pad when $\mu_0H=129$ mT and for the wire with $\mu_0H=198$ mT.
In Fig.~\ref{fig4} (a), the real-space distribution of the time averaged precession angle $\bar{\theta}$ is shown at the FMR of the wire with $\alpha=0.003$.
Here $\bar{\theta}=1/T\int_0^T dt\theta(t)$, where $T=1/f$ is the period of precession.
Because of the demagnetizing field, the magnetization precession is strongly pinned near the sample edge, even at resonance.
In the corners of the sample, the magnetization is pinned at a large  angle to the magnetic field.
In the wire, the magnetization is strongly affected by this pinning effect, resulting, with $\alpha=0.003$, in the small precession angle $5.5^\circ$ at the wire resonance, see Fig.~\ref{fig4} (a), as compared with $9.8^\circ$ for the pad (not shown).
As a result, the motive force is larger for the pad resonance.

The spinmotive force $V(t)$ is defined by the spatial difference of the electric potential $\phi\left(\bm{r},t\right)$, which is obtained by solving the Poisson equation $\nabla^2\phi=-\nabla\cdot\bm{E}_\mathrm{s}$\cite{yang,ohe}, as $V(t)=\phi(\bm{r}_W,t)-\phi(\bm{r}_P,t)$.
Here, $\bm{r}_P$ $(\bm{r}_W)$ indicates the middle of the pad (wire), see Fig.~\ref{fig4} (a), corresponding to the experimental situation where the electrodes are attached to the sample as shown in Fig.~\ref{fig1} (c).
Figure \ref{fig4} (b) shows the time-averaged spinmotive force $\bar{V}$ as a function of the external static magnetic field $H$ for $\alpha=0.01$, $0.006$ and $0.003$, with a definition of $\bar{V}$ analogous to that of  $\bar{\theta}$.
A dc electromotive force appears at each ferromagnetic resonance.
The spinmotive force is larger for the smaller $\alpha$ due to the increased precessional angle, and it is found that when $\alpha=0.003$ the calculated values reproduce well the experimental data.
The spinmotive force is calculated with different ac magnetic field amplitudes $h$, i.e., for different microwave power.
The results are shown in Fig.~\ref{fig3}.

The inset of Fig.~\ref{fig4} (b) compares the analytic result Eq.~(\ref{v}), the grey curve, with the numerical results on resonance with $\alpha=0.003$.
The red dot corresponds to the pad and green equivalent to the wire.
The horizontal dotted lines reflect the experimental results for which the angle is not accessible.
While the theory values are in reasonable agreement for the wire the result of the numerical calculations falls below the analytic result for the pad.
In this context, it is noted that the precession trajectory of the pad megnetization is more elliptic than in the wire because of the larger out-of-plane anisotropy in the former case and that $\mu_0H=129$ mT is insufficient to align, against the shape anisotropy, the wire magnetization along the $z$-axis.
The static magnetization near the center of the wire makes an angle $\bar{\theta}\sim 22^\circ$ with $\bm{H}$.
These facts are not accounted for in Eq.~(\ref{v}) and indicate the utility of the numerical methods.

%===================================================================================================
%           Discussion
%===================================================================================================
\emph{Discussion.}---
A dc voltage is generated by exciting the FMR in a lateral F/N junction\cite{pump1} via ``spin-pumping"\cite{tser1}, which, at first sight, would seem to be similar to the present configuration but with the N-layer replaced by the non-resonating part of our single ferromagnetic layer.
In their scheme, the voltage is due to the spin accumulation at the relevant interface, here that between the two parts of the ferromagnet with the actual metal contacts being irrelevant since the voltage is developed within the magnet.
More recently, a large voltage of several $\mu$V was observed in a F/insulator(I)/N multilayer\cite{pump2}, and a F/I/F junction has been studied theoretically\cite{tser2}.
In contrast, for the present patterned single sub-millimeter Permalloy sample, within which there is no well defined interface, there is a negligibly small spin accumulation around the junction between the pad and wire making it implausible that the present experiment can be explained in terms of ``spin pumping". 

%Finally, the thermoelectric effect is not expected to contribute to the observed electromotive force since such a sub-millimeter metallic sample will reach a uniform thermal equilibrium state for the present experimental conditions\cite{thermo}.

Finally, let us discuss the thermoelectric effect.
The present metallic sample of sub-millimeter scale will reach a thermal equilibrium state very quickly\cite{thermo}.
Furthermore, the Seebeck coefficient of the electrodes (Au$_{25}$Pd$_{75}$), $-22 \mu$V/K\cite{aupd}, and of the sample (Ni$_{81}$Fe$_{19}$), $-20 \mu$V/K\cite{nife}, are almost the same.
This fact results in very small thermoelectric voltage even if there exists finite temperature difference in the sample.
For these reasons, we conclude that the thermoelectric effect does not contribute to the observed electromotive force.

%===================================================================================================
%           Summary
%===================================================================================================
\emph{Summary.}---
We have produced a continuous dc spinmotive force by the resonant microwave excitation of a ferromagnetic ``comb" structure patterned from a uniform ferromagnetic thin film.
The experimental results are fully consistent with the theoretically predictions for such a motive force, lending strong support for this concept.

%===================================================================================================
%           Acknowledgments
%===================================================================================================
We are grateful to K.~Uchida, Y.~Kajiwara, K.~Ando of Tohoku University for valuable discussions, and H.~Adachi, Japan Atomic Energy Agency, for helpful comments on this work.
This research was supported by a Grant-in-Aid for Scientific Research from MEXT, Japan and the Next Generation Supercomputer Project, Nanoscience Program from MEXT, Japan.

%===================================================================================================
%           References
%===================================================================================================


\begin{thebibliography}{99}

%\bibitem{barnes1} S. E. Barnes and S. Maekawa, Phys. Rev. Lett. {\bf{98}}, 246601 (2007).
%\bibitem{yang}    S. A. Yang, G. S. D. Beach, C. Knutson, D. Xiao, Q. Niu, M. Tsoi, and J. L. Erskine, Phys. Rev. Lett. {\bf{102}}, 067201 (2009); 
%                  S. A. Yang, G. S. D. Beach, C. Knutson, D. Xiao, Z. Zhang, M. Tsoi, Q. Niu, A. H. MacDonald,  and J. L. Erskine, Phys. Rev. B, {\bf{82}}, 054410 (2010).
%\bibitem{tanaka}  P. N. Hai, S. Ohya, M. Tanaka, S. E. Barnes, and S. Maekawa, Nature {\bf{458}}, 489 (2009).
%\bibitem{barnes2} S. E. Barnes, J. Ieda, and S. Maekawa, Appl. Phys. Lett. {\bf{89}}, 122507 (2006).
%\bibitem{field}   G. E. Volovik, J. Phys. C {\bf{20}}, L83 (1987); 
%                  M. Stamenova, T. N. Todorov, and S. Sanvito, Phys. Rev. B {\bf{77}}, 054439 (2008); 
%                  R. A. Duine, Phys. Rev. B {\bf{77}}, 014409 (2008); 
%                  Y. Tserkovnyak and M. Mecklenburg, Phys. Rev. B {\bf{77}}, 134407 (2008); 
%                  Y. Yamane, J. Ieda, J. Ohe, S. E. Barnes, and S. Maekawa, J. Appl. Phys. {\bf{109}}, 07C735 (2011).
%\bibitem{ohe}     J. Ohe and S. Maekawa, J. Appl. Phys. {\bf{105}}, 07C706 (2009); 
%                  J. Ohe, S. E. Barnes, H. W. Lee, and S. Maekawa, Appl. Phys. Lett. {\bf{95}}, 123110 (2009).
%\bibitem{pump1}   X. Wang, G. E. W. Bauer, B. J. van Wees, A. Brataas, and Y. Tserkovnyak, Phys. Rev. Lett. {\bf{97}}, 216602 (2006); 
%                  M. V. Costache, M. Sladkov, S. M. Watts, C. H. van der Wal, and B. J. van Wees, Phys. Rev. Lett. {\bf{97}}, 216603 (2006).
%\bibitem{oommf}   http://math.nist.gov/oommf/
%\bibitem{fmr}     H. M. Olson, P. Krivosik, K. Srinivasan, and C. E. Patton, J. Appl. Phys. {\bf{102}}, 023904 (2007).
%\bibitem{tser1}   Y. Tserkovnyak, A. Brataas,  and G. E. W. Bauer, Phys. Rev. Lett. {\bf{88}}, 117601 (2002).
%\bibitem{pump2}   T. Moriyama, R. Cao, X. Fan, G. Xuan, B. K. Nikoli\'{c}, Y. Tserkovnyak, J. Kolodzey, J. Q. Xiao, Phys. Rev. Lett. {\bf{100}}, 067602 (2008).
%\bibitem{tser2}   Y. Tserkovnyak, T. Moriyama, and J. Q. Xiao, Phys. Rev. B {\bf{78}}, 020401(R) (2008).
%\bibitem{thermo}  The time required for the equilibrium temperature to be achieved throughout the sample is estimated as $\tau\sim l^2\rho c_p/\kappa\sim  10^{-6}$ s [L. D. Landau and E. M. Lifshitz, \textit{Fluid Mechanics, Second Edition, Chapter} V, (1987)], which is much shorter than the time scale of the lock-in modulation.
%                  Here $l^2\sim 10^{-11}$ m$^2$ is the relevant area of the present sample, $\rho=8.7\times 10^3$ kg/m$^3$ is the electron density, $c_p=4.2\times 10^2$ J/kg/K is the specific heat, and $\kappa=3.4\times 10$ J/m/K/s is the heat conductivity, respectively.


\bibitem{barnes1} S. E. Barnes and S. Maekawa, Phys. Rev. Lett. {\bf{98}}, 246601 (2007).
\bibitem{yang}    S. A. Yang \textit{et al.}, Phys. Rev. Lett. {\bf{102}}, 067201 (2009); 
                  S. A. Yang \textit{et al.}, Phys. Rev. B, {\bf{82}}, 054410 (2010).
\bibitem{tanaka}  P. N. Hai \textit{et al.}, Nature {\bf{458}}, 489 (2009).
\bibitem{barnes2} S. E. Barnes \textit{et al.}, Appl. Phys. Lett. {\bf{89}}, 122507 (2006).
\bibitem{field}   G. E. Volovik, J. Phys. C {\bf{20}}, L83 (1987); 
                  M. Stamenova \textit{et al.}, Phys. Rev. B {\bf{77}}, 054439 (2008); 
                  R. A. Duine, Phys. Rev. B {\bf{77}}, 014409 (2008); 
                  Y. Tserkovnyak and M. Mecklenburg, Phys. Rev. B {\bf{77}}, 134407 (2008); 
                  Y. Yamane \textit{et al.}, J. Appl. Phys. {\bf{109}}, 07C735 (2011).
\bibitem{ohe}     J. Ohe and S. Maekawa, J. Appl. Phys. {\bf{105}}, 07C706 (2009); 
                  J. Ohe \textit{et al.}, Appl. Phys. Lett. {\bf{95}}, 123110 (2009).
\bibitem{pump1}   X. Wang \textit{et al.}, Phys. Rev. Lett. {\bf{97}}, 216602 (2006); 
                  M. V. Costache \textit{et al.}, Phys. Rev. Lett. {\bf{97}}, 216603 (2006).
\bibitem{oommf}   http://math.nist.gov/oommf/
\bibitem{fmr}     H. M. Olson \textit{et al.}, J. Appl. Phys. {\bf{102}}, 023904 (2007).
\bibitem{tser1}   Y. Tserkovnyak \textit{et al.}, Phys. Rev. Lett. {\bf{88}}, 117601 (2002).
\bibitem{pump2}   T. Moriyama \textit{et al.}, Phys. Rev. Lett. {\bf{100}}, 067602 (2008).
\bibitem{tser2}   Y. Tserkovnyak \textit{et al.}, Phys. Rev. B {\bf{78}}, 020401(R) (2008).
\bibitem{thermo}  The time required for the equilibrium temperature to be achieved throughout the sample is estimated as $\tau\sim l^2\rho c_p/\kappa\sim  10^{-6}$ s [L. D. Landau and E. M. Lifshitz, \textit{Fluid Mechanics, Second Edition, Chapter} V, (1987)], which is much shorter than the time scale of the observation.
                  Here $l^2\sim 10^{-11}$ m$^2$ is the relevant area of the present sample, $\rho=8.7\times 10^3$ kg/m$^3$ is the electron density, $c_p=4.2\times 10^2$ J/kg/K is the specific heat, and $\kappa=3.4\times 10$ J/m/K/s is the heat conductivity, respectively.
\bibitem{aupd}    T. Rowland \textit{et al.}, J. Phys. F: Met. Phys. {\bf{4}}, 2189 (1974)
\bibitem{nife}    K. Uchida \textit{et al.}, Nature (London) {\bf{455}}, 778 (2008)
\end{thebibliography}
\end{document}